\documentclass[onecolumn]{article}
\usepackage{amsmath,amssymb}

\input xy
\xyoption{all}
\xyoption{arc}
\xyoption{line}
\xyoption{rotate}
\xyoption{color}
\usepackage[curve]{xy}
\usepackage{hyperref}
\begin{document}
\title{Some Unrecorded Science History\\
My Brush with Einstein's Relativistic Length Contraction}
\author{James G. Gilson\quad  j.g.gilson@qmul.ac.uk\thanks{
 School of Mathematical Sciences,
Queen Mary University of London, Mile End Road, London E1 4NS,
United Kingdom.}}
\date{February 22, 2007}

\maketitle

\begin{abstract}
This paper is a short historical note about the origin of the {\it Pole-Barn\/} and the {\it Fat Man\/} Paradoxes.
\end{abstract}
\section{Introduction}
\setcounter{equation}{0}
\label{sec-intr}
I first became interested in Einstein's relativity theory in pre-war days after coming across a copy of his famous book {\it The Meaning of Relativity\/} in the local library. I barely understood it, but was set off on the idea and determination that one day I would really understand what it was all about. I was called up for service in the Royal Navy and demobbed after the war and was lucky to get a further education grant, available to ex-service men, to attend university. My next lucky break was to find myself in a university department where Wolfgang Rindler was teaching relativity. His lectures were truly memorable for their clarity and comprehensiveness.  On one occasion, he was talking at length about the Einstein relativistic length contraction that Einstein predicted should result from the velocity of a physical body in motion relative to the observer of the motion. This was one of Einstein's revolutionary new ideas, seemingly quite nonsensical in terms of the earlier {\it common sense\/} of the Newtonian concepts of a rigid body in motion. How could a rigid body contract?  This question was a philosophical dilemma for the scientific community many of whom were inclined to accept the mathematics of Einstein's ideas but could only go along with the contraction idea by adding the caveat that the contraction was not real but rather an illusion resulting from the motion and the way the observation was made. Wolfgang was firmly convinced that the result was real and therefore was a technically usable consequence of relativity. Another lecturer on relativity in the physics department of the same institution took exactly the opposite view, that the contraction was not real. There was at that time, the fifties, totally conflicting ideas of the philosophical significance of the length contraction idea.  I was also strongly convinced that length contraction was a reality because the contraction with velocity factor involved $(1 - (v/c)^2)^{1/2}$ occurred in so very many situations in relativistic physical theory that were verifiable that it seemed to me that it had to be accepted as real. However, I also had in mind a graphical proof that seemed to show that under some conditions the contraction effect could not be actually used in the practical context. Thus it was that I proposed the following {\it paradox\/} to Wolfgang Rindler during one of his lectures on relativity.

Suppose a {\it fat\/} man is running fast with a speed $v$ over a drain grating in a direction perpendicular to the {\it narrow slits\/} in the grating. The fat man, as the observer according to relativity,  will see the narrow slits as even narrower because they are passing him with velocity $-v$ in the opposite direction to that which he is  running relative to the road and so he will be confident that he cannot fall through the grating. However another observer standing by the drain cover or in the road will, according to relativity, see a {\it thin\/} man running past and will fear for the runners safety. This very simple but highly ambiguous observational situation derivable from special relativity theory, I called the {\it Fat Man Paradox\/} \footnote{Professor Rindler had used the fat runner falling through grating image in relation to the reality issue before my intervention on the symmetry issue}. 
Wolfgang was unsure of the answer but because he believed in the {\it reality\/} of the relativity contraction idea, he thought the fat man would fall into the drain. After this lecture we had a tutorial with Wolfgang. I was having an argument with a student sitting next to me about the paradox who also thought the runner could fall down the drain and in the course of this argument I made the remark {\it Rindler is talking through his hat\/} as I had concluded the runner could not fall into the drain. Unfortunately, I had not noticed Wolfgang was sitting in the next row directly in front of me giving tutorial help to another student. To this day, I do not know if Wolfgang heard this remark but certainly I was very embarrassed by my typical student irreverence. If you are reading this Wolfgang, I apologise for my rudeness.

Wolfgang went to America and became recognised as a leading World Cosmologist as a result of his research on relativity. He also wrote a number of remarkable books  of great clarity and originality based on his fifties lectures.
His most recent book \cite{03:rind}, {\it Relativity: Special, General
and Cosmological, Second Edition, Oxford University Press\/}, is the one mostly relevant to this article.
In 1961, Wolfang wrote a paper on the {\it Fat Man paradox\/} and published it in the American Journal of Physics (\cite{42:rind}). This essentially showed the reality of length contraction. He had, I understand, taken about eight years pondering this paradox before coming to any conclusions. Meanwhile, I had taken up university teaching and in $1964$ joined the Mathematic staff at Queen Mary College London and found myself teaching relativity for which I used one of Wolfgang's books for course work. During my lectures I used the fat man paradox to illustrate length contraction rather as a little light entertainment in the lectures. After one such lecture I was waylaid in a corridor by a very irate colleague who was teaching some relativity in connection with another course. He suggested that I had leaked one of his exam questions about the {\it Fat Man\/} paradox to his students. He gave me the strong impression that he thought I was stealing his ideas. I explained that this was part of my course structure and contents which he seemed to reluctantly accept. I did not tell him that I was the originator of the idea. It does, however, show how widespread this paradox had permeated relativity teaching while his reaction seemed something of a joke at the time. 

A few years later, I came to realise that the length contraction effect was central to developing a formula for the {\it fine structure constant\/} \cite{32:gil}. I wrote my first paper on that subject in $1994$ and sent a copy of that paper to Wolfgang Rindler whom I had not seen or communicated with for about $38$ Years.
He replied and said he remembered me very well and thanked me for my input of the paradox for his $1961$ paper \cite{42:rind} on the subject. Also see footnote page $63$ of Rinder's book \cite{03:rind}. In the next section, I shall give the case with a figure and explanation as to why I did not think the fat man could fall down the drain.

The Barn-Pole paradox is a derivative of the Fat Man paradox but the former has some different facets from the later which make the questions that one can ask about them not quite equivalent. Wolfgang mentions them together in his latest book and perhaps does not give sufficient emphasis to their distinctiveness. The barn pole paradox which also starts off as a three dimensional idea but is in essence one  dimensional in contrast with the fat man paradox which starts off as a three dimension idea but is in essence two dimensional. They both depend essentially on the relativistic length contraction idea.
\section{Can the running Fat Man fall down the Drain?}
The diagram below is what I had in mind when I made the regrettable and impolite remark about Wolfgang. The diagram represents the view of the situation of an observer standing by the drain opening, its short dimension is the hole in the road between the two {\it Hits\/} positions. The long dimension of the grating is perpendicular to the page and so is not shown but is assumed to be greater than the diameter of the sphere of which the outer circle is a great circle. The outer circle can also be taken to represent the stationary observer standing by the drain. Thus the stationary man according to relativity sees the inner vertical ellipse shrunk from the circular shape because of its centroidal, horizontal to the left, speed component $v$. Thus the outer circle besides representing the fixed observer can also be taken to represent the rest frame shape of the runner. It is clear from this why the stationary man fears that the runner can fall through the grating as his thin man appearance can be tracked down to the {\it Clear, Clear\/} positions  missing the drain edges if he should fall with the additional downwards velocity component $v_d$. As we have noticed earlier a diagram from the point of view of the runner would only show a further reduced drain width opening due to its velocity $-v$  to the right so that from his point of view there is no problem. He is convinced that he could not fall through such a narrow opening.  Thus we have to reconcile these two apparently contradictory points of view. From the diagram this is easy, if we take into account the true significance of {\it relativistic length contraction with speed\/} which should be expressed as:

\centerline{\xy
\POS(-150,0)*+{
\xybox{
\POS(-28.5,53)*+{\phantom{Z}}                \POS(5,67.5)*+{\xybox{,(3,100)*+{\phantom{A}};(70,100)*+{\phantom{B}};0;/r5pc/:*\dir{},*++!DR(.1){};
p+(.5,-.1)*\dir{}="c"
**\dir{},*+!UL{},"c"
,{\ellipse<40pt,20pt>{--}}}}
\POS(0,105)*+{\it\ The\ relativistic\ length\ contraction\ of\ an}
\POS(0,95)*+{\it\ object\ with\ speed\ only\ involves\ the\ dimension\ of\/}
\POS(0,85)*+{\it\ the\ object\ lying\ in\ the\ direction\ of\ motion\/}.    
\POS(0,60)*+{\bf {Fat\ Man\ Paradox\ Diagram}}
\POS(0,45)*+{What\ the\  man\ standing\ by\ the\ grating\ sees}
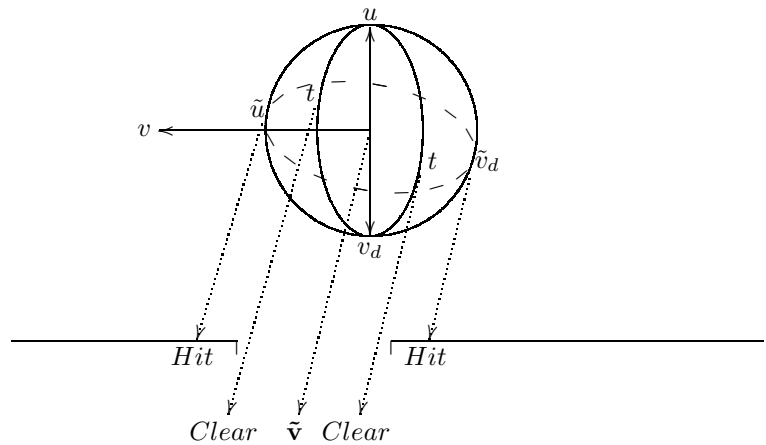
\POS(0,20)*+{}
\ar @{->} (0,35.5)*+{u}
\ar @{->} (-30,20)*+{v}
\ar @{->} (0,4)*+{v_d}
\ar @{.>} (-10,-20)*+{\bf \tilde v}
\ar @{} (-33,+25.6)*+{}
\ellipse<40pt,20pt>{-}
\POS(-1,+20)
\ellipse<40pt,40pt>{-}
\POS(-16.1,23)*+{\tilde u\ \ }
\ar @{.>} (-26,-10)*+{Hit}
\POS(+11.5,16)*+{\ \ \ \tilde v_d}
\ar @{.>} (5,-10)*+{Hit}
\POS(-50,-8)*+{}
\ar @_{|} (-20,-8)*+{}
\POS(+50,-8)*+{}
\ar @^{|} (0.5,-8)*+{}
\POS(-9.1,25)*+{t\ \ }
\ar @{.>} (-21.8,-20)*+{Clear}
\POS(+4.8,16)*+{\ \ t}
\ar @{.>} (-4.2,-20)*+{Clear}}}
\endxy}
\vskip 1cm

Inspecting the diagram reveals that if the thin looking runner is to faultlessly fall through the hole he must acquire the  extra downwards speed $v_d$ resulting in a {\it resultant\/} velocity ${\bf \tilde v}$ that has a direction so that his centroid passes through the centroid of the hole. This would be OK, if his shape orientation remained as the vertical  ellipse $v_d,t,u,t$. However his orientation  does not remain the same because his resultant velocity will have changed from $v$ to ${\bf \tilde v}$ and the moving ellipse's major axis will have changed direction from vertical to $(\tilde u, \tilde v_d)$. Thus the falling ellipse will have the shape orientation of the dashed ellipse in the diagram. Consequently the fall causes bad injuries at both his ends at the {\it Hit, Hit\/} moment. It seems that according to Einstein's special relativity a {\it sphere cannot\/} pass through a rectangular hole if its diameter is greater than the smallest side of the rectangular hole. This diagrammatic argument is not a rigorous proof of the previous statement but I think it could be made so.
\section{Conclusions}
I shall come to no technical conclusions here but instead make a few remarks about the fat man paradox. Gravity need not be taken into account in this puzzle for the following reason. Suppose the stationary observer is a baddy and that there is no gravity present. He decides to settle the issue by hitting a passing fat man on the head. Referring to the diagram again, we see this will induce the required downward velocity $v_d$ and by the diagram any conclusions will remain unaffected. My intention in writing this article was only to fill in a little unrecorded science history. However, it might stimulate some further interest in the philosophical issue of what reality in physical theory means. It does not seem always to mean {\it usability\/}!

\vskip 0.5cm
\leftline{\bf Acknowledgment}
\vskip 0.3cm
\leftline{I am greatly indebted to Professor Wolfgang Rindler for those} \leftline{remarkably clear and inspirational lectures many years ago.}

\end{document}